\documentclass[pdftex]{article}
\usepackage{icrctc07}

\title{Observations of Mrk~421 and Mrk~501 in Spring 2006 with VERITAS}
\shorttitle{Observations of Mrk~421 and 501 with VERITAS}
\authors{S. J. Fegan$^{1}$ for The VERITAS Collaboration$^{2}$.}
\shortauthors{S. J. Fegan and et al}
\afiliations{$^1$Division of Physics and Astronomy, University of California at Los Angeles, 430 Portola Plaza, Box 951547, Los Angeles, CA 90095-1547 \\
$^2$ For full author list see G. Maier, ``Status and Performance of VERITAS'',
these proceedings.}
\email{sfegan@astro.ucla.edu}

\abstract{VERITAS, the Very Energetic Radiation Telescope Imaging Array 
System, is an array of four imaging Atmospheric Cherenkov telescopes
in southern Arizona. It is sensitive to gamma rays at energies above
100GeV. Here, we discuss the results of observations of two well known
VHE blazars, Markarian 421 and Markarian 501, during Spring 2006 which
were made with the first two telescopes during the comissioning phase
of VERITAS. During most of this time Mrk~421 was in an unusually active
state while Mrk~501 was in a much lower flux state. As such, these
observations provided an opportunity to test the sensitivity of the
instrument to strong and weak sources.  We discuss implications of
these observations on our understanding of these objects.}

\begin{document}
\maketitle

\section{Introduction}

The blazars Mrk~421 and Mrk~501 were the first extragalactic sources
from which very high energy (VHE; E$>$200GeV) gamma-ray emission was
definitively observed, when they were detected with the Whipple 10m
telescope over a decade ago \cite{REF::PUNCH::NATURE1992,
REF::QUINN::APJ1996}. They undergo periods of ``flaring'',
characterized by rapid changes in the flux of VHE gamma rays
\cite{REF::GAIDOS::NATURE1996}, often correlated with an increased flux 
at hard X-ray energies \cite{REF::MARASCHI::AP1999}. During flaring
episodes the flux from these objects has been observed to increase by
over a factor of ten, making them the brightest sources in the sky in
the VHE regime. Flux doubling time scales as short as a few minutes
and changes in the shape of the spectrum of VHE gamma rays have also
been observed \cite{REF::GAIDOS::NATURE1996, REF::KRENNRICH::APJ2002}.
During the long quiescent periods between flares, often lasting many
months, the flux from Mrk~501 was too low to be detected by the
previous generation of VHE instruments, such as the Whipple 10m and
HEGRA array \cite{REF::CAWLEY::ICRC1990,REF::AHARONIAN::ECRS1997}.

Blazars are understood to be members of the broader class of Active
Galactic Nuclei (AGN), galaxies in which broadband emission from the
central core region dominates the combined luminosity of the billions
of stars. Emission from the core of AGN is thought to be powered by
accretion onto a supermassive black hole ($\sim10^8M_\odot$). VHE
gamma rays seem to be associated with a jet of high-energy particles
accelerated close to the black hole, which emanates from the core in a
direction perpendicular to plane of accretion. Jets have been observed
from many AGN at radio wavelengths. In blazars, the high-energy
particles in the jet are thought to be moving directly along our
line-of-sight to the AGN, and hence the jet cannot be resolved
directly. The VHE gamma rays associated with the flaring are usually
explained as arising through inverse-Compton scattering of low-energy
photons by relativistic electrons in the jet. These photons could come
from the environment external to the jet or through processes in the
jet itself, such as synchrotron emission. The role of protons, which
must be present in the jet at some level, is unclear. The fast
acceleration and cooling times associated with the flaring is
indicative of electrons, the slower timescales associated with protons
may mean that VHE gamma-ray emission through hadronic processes
contributes to the more steady, baseline emission from blazars.
Measuring the emission from blazars in all flux states is important to
understand the details of relativistic particle populations in the jet.

The VERITAS gamma-ray observatory, an array of four 12m atmospheric
Cherenkov imaging telescopes in southern Arizona,
U.S.A. \cite{REF::WEEKES::AP2002,REF::MAIER::ICRC2007_STATUS},
represents a significant increase in sensitivity over the previous
generation of instruments operating in the VHE regime. This
sensitivity is achieved through two primary innovations: large
telescopes to collect Cherenkov photons from distant air showers, and
stereoscope imaging, in which the development of air showers in the
atmosphere is measured from multiple locations on the ground, allowing
the properties of the primary, its species, direction, impact point
and energy, to be determined. The VERITAS telescopes are separated by
approximately 85-100m on the ground, optimized to give good response
in the 100-200\,GeV energy range. Each camera consists of 499
photo-multiplier tubes and a flash-ADC-based data acquisition system,
through which the shower development is digitized with a resolution of
0.15$^\circ$ in the angular domain and 2~ns in the time
domain. Readout is initiated by a three level trigger system, which
detects potentially interesting signals in individual channels (level
1, or L1), in individual telescopes (L2) or in the full array
(L3). Only when two telescopes detect a coincident signal in at least
three neighboring camera pixels does the system record an image. These
images are stored to disk, along with housekeeping information,
including telescope pointing information, rates of triggering of each
channel and each telescope, sky temperature, and so on. The events are
processed off line using multiple independent analysis packages, see
\cite{REF::DANIEL::ICRC07_ANALYSIS} for example. These analyses yield
consistent results.

Observations of AGN, to measure the properties of the populations of
relativistic particles and to reveal the nature of the mechanism
responsible for their acceleration, is one of the primary scientific
goals of VERITAS. This paper presents the results from the first
observations of AGN with VERITAS.

\section{Observations}

Mrk~421 and Mrk~501 were observed as part of the process of
commissioning VERITAS. The second telescope and the array trigger,
which allows the telescopes to operate together, were installed during
Spring 2006. A total of $\sim17$ hours of observations were made of
Mrk~421 in April 2006, and $\sim18.5$ hours of Mrk~501 between April
and June. Initially all data were taken in ``pairs'' mode, in which
each half hour observation is matched with a dedicated ``off source''
observation which is used for the purposes of background
estimation. Although this mode represents an inefficient use of the
telescope time, the independent background estimate is useful for
understanding the performance of the newly constructed
instrument. During the spring of 2006, VERITAS transitioned to making
almost all observations in ``wobble'' mode, in which the putative
source is always in the field of view of the instrument, offset from
the center, so as to allow a simultaneous measurement of the
background. Table~\ref{TAB::DATA} gives the details of the VERITAS
observations of the two blazars.

\begin{table*}[ht]
\caption{Details of observation presented in this paper. Earlier
observations, from April, were made in a mixture of Pairs and Tracking
modes. Later observations were largely in Wobble mode.}
\label{TAB::DATA}
\begin{center}
\begin{tabular}{lllll}\hline
\textbf{Target} & \textbf{Period} & \textbf{Pairs [hr]} & \textbf{Wobble [hrs]} & \textbf{Tracking [hrs]}\\\hline
Mrk~421 & 04/06          & 4.5 &  5.0 & 7.5 \\ 
Mrk~501 & 04/06 -- 06/06 & 1.5 & 12.5 & 4.5 \\\hline
\end{tabular}
\end{center}
\end{table*}

\section {Results}

\begin{figure*}[t]
\begin{center}
\resizebox{0.85\textwidth}{!}{\includegraphics[height=0.5\textwidth]{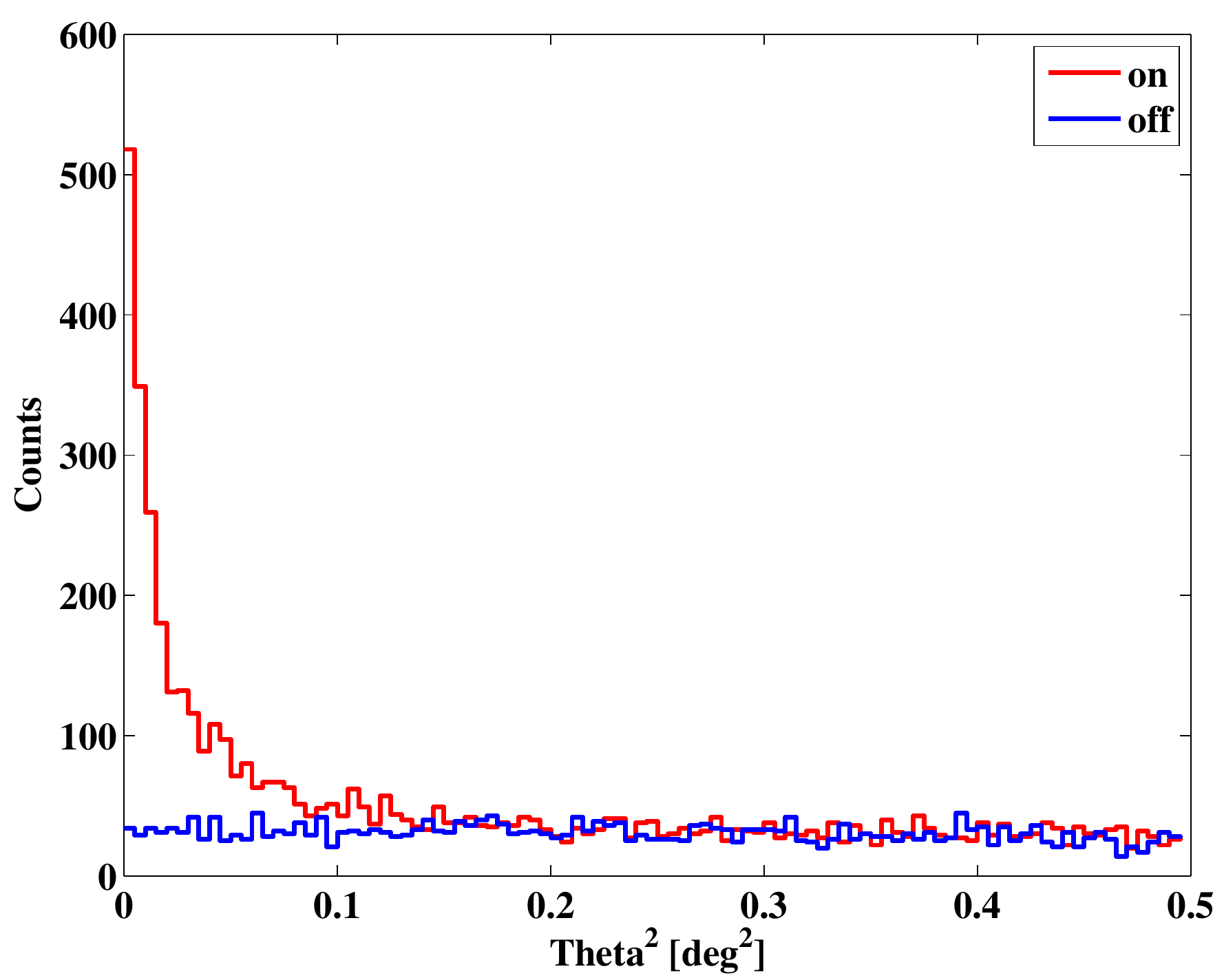}%
\includegraphics[height=0.5\textwidth]{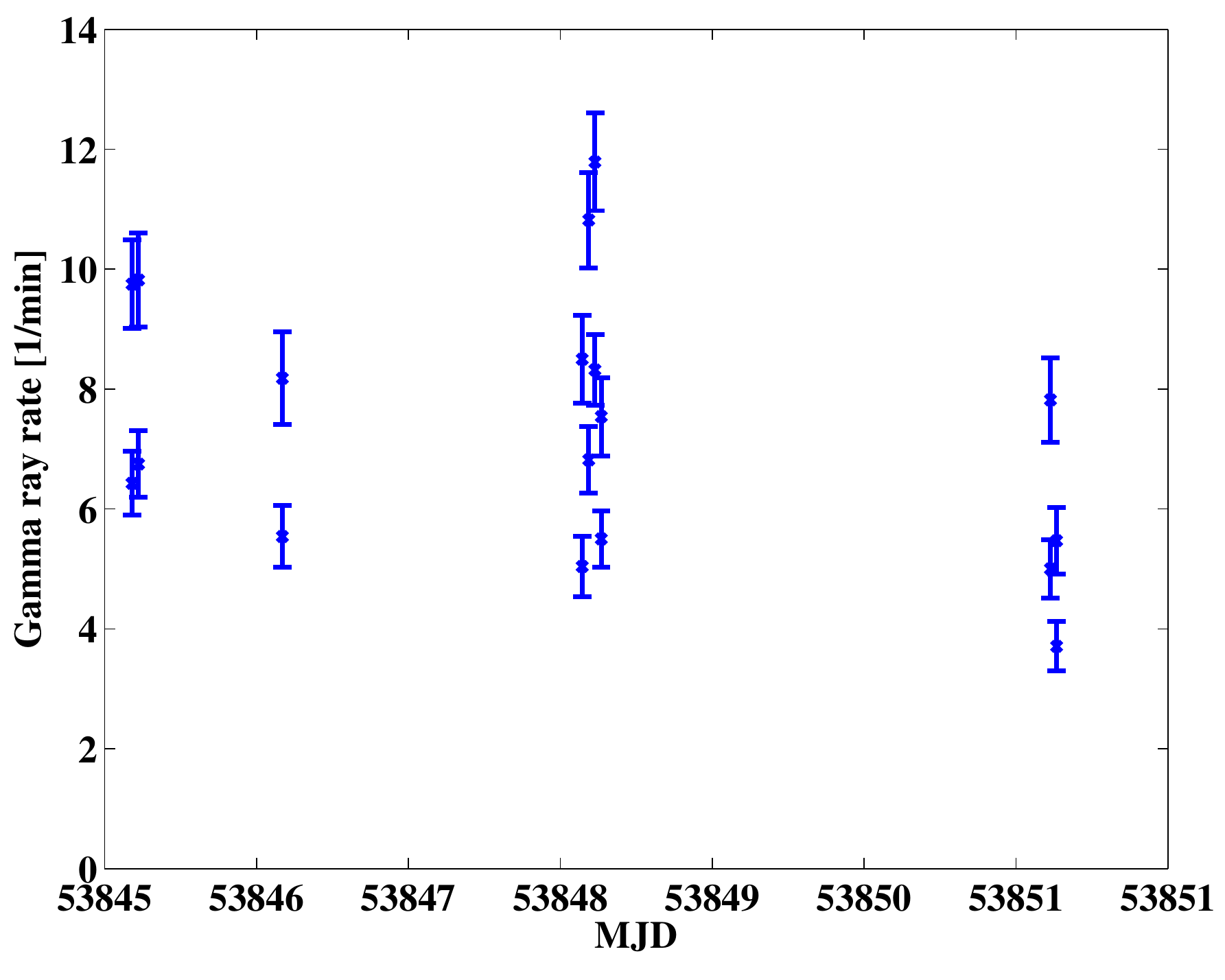}}
\end{center}
\caption{Left: number of gamma-ray-like events as a function of
squared angular distance between the reconstructed gamma-ray 
origin and the target,
Mrk~421 (curve marked as ``ON''). Also shown is the number from the
background observation, in the absence of a source (marked as
``OFF''). Right: rate of gamma rays detected from Mrk~421 with
VERITAS during April 2006.}\label{FIG::MRK421}
\end{figure*}

During the period of the observations, Mrk~421 was in a particularly
active state and was consistently detected by VERITAS and, at lower
significance, by the Whipple 10m
instrument. Figure~\ref{FIG::MRK421}~(left) shows the density of
gamma-ray-like events from the 4.5 hours of Mrk~421 ``pairs''
observations (and for the dedicated background observation), as a
function of $\theta^2$, the squared distance between the target and
the reconstructed gamma ray. The clear excess at small values of
$\theta^2$ is consistent with emission from Mrk~421 at the 35$\sigma$
level, with a rate of 5.6 gamma rays/minute after the data selection
criteria are applied. Numerous analyses by various groups within the
VERITAS collaboration give similar results using different analysis
codes and reconstruction methodologies.

\begin{figure}[ht]
\begin{center}
\includegraphics[width=0.38\textwidth]{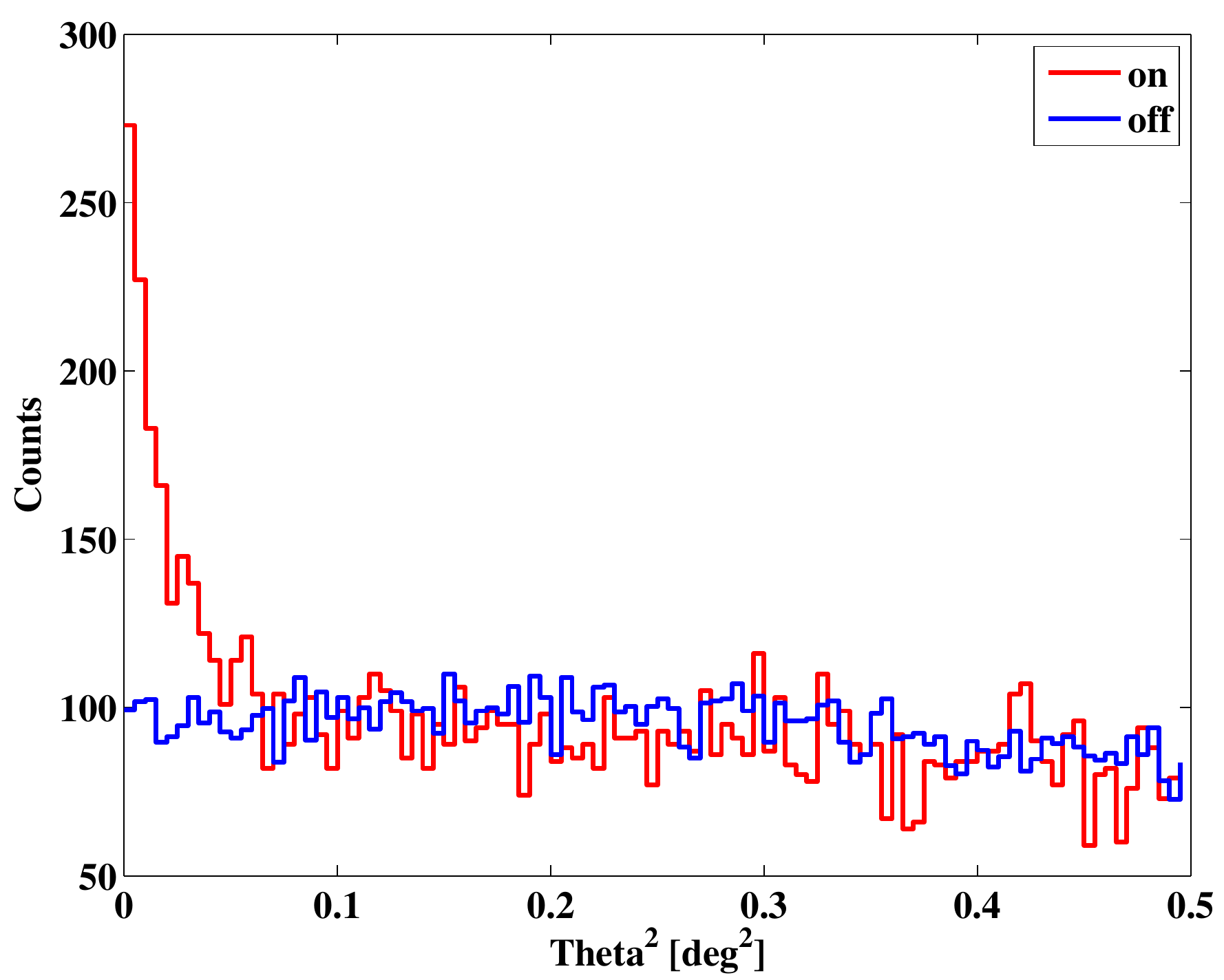}
\end{center}
\caption{Number of of gamma-ray like events for 12.5 hours of
Mrk~501 ``wobble'' observations. A clear excess can be seen,
consistent with gamma-ray emission from Mrk~501.}\label{FIG::MRK501THETA2}
\end{figure}

The results of the 12.5 hours of wobble mode observations of Mrk~501
are shown in figure~\ref{FIG::MRK501THETA2}. In this mode the signal
and background are estimated from the same data set. The `ON'' curve
shows shows the $\theta^2$ parameter for each event measured from the
target position. The ``OFF'' curve shows the same parameter from three
different background positions in the sky (summed into one histogram
and divided by three). Therefore each event appears in the plot four
times. However, by construction, events can appear only one time in
the region $\theta^2<0.035$, from which the signal and background are
estimated. The results indicate a $16\sigma$ detection, with a mean
rate of 0.8 gamma rays/min.

\begin{figure*}
\begin{center}
\resizebox{0.85\textwidth}{!}{\includegraphics[height=0.5\textwidth]{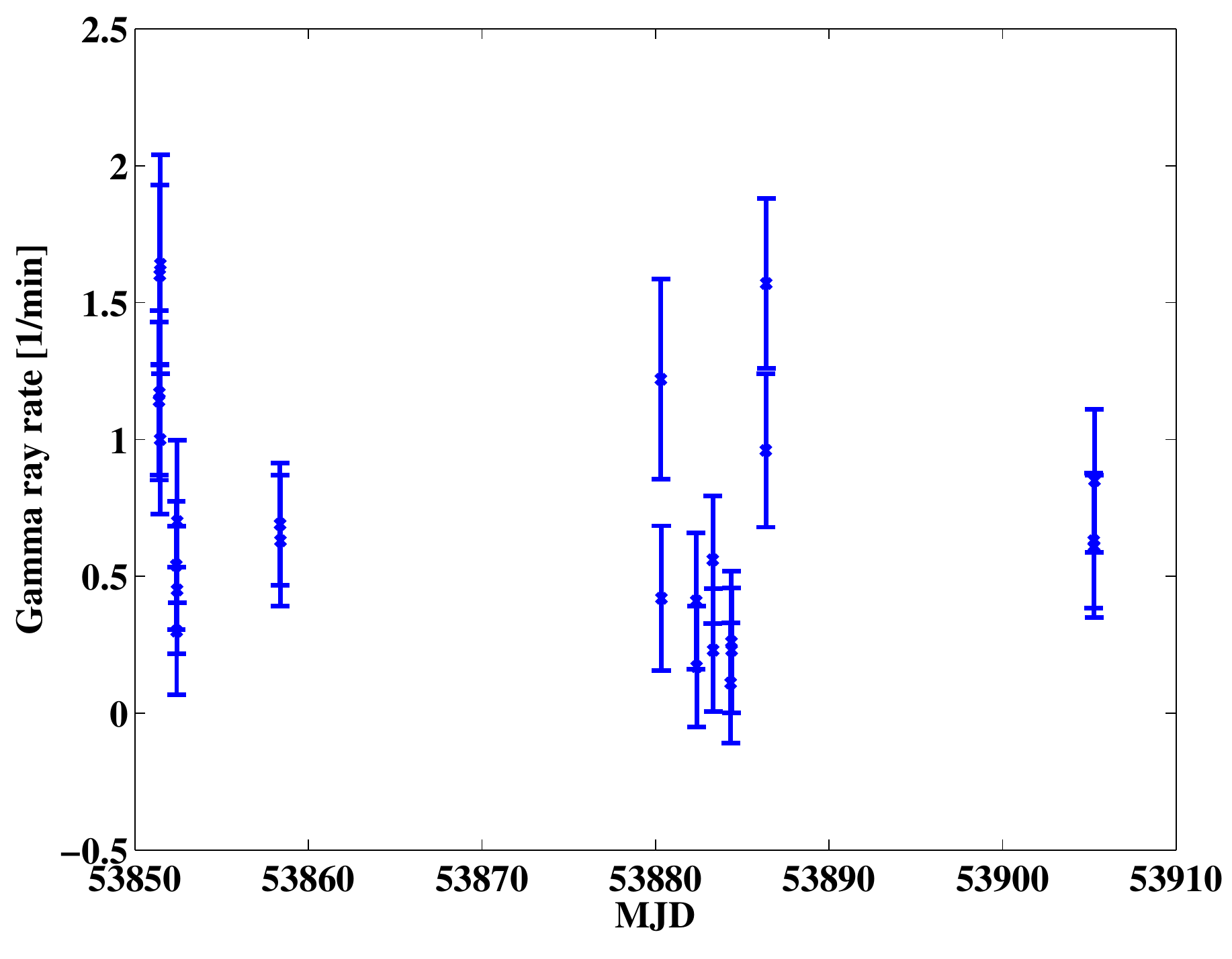}%
\includegraphics[height=0.5\textwidth]{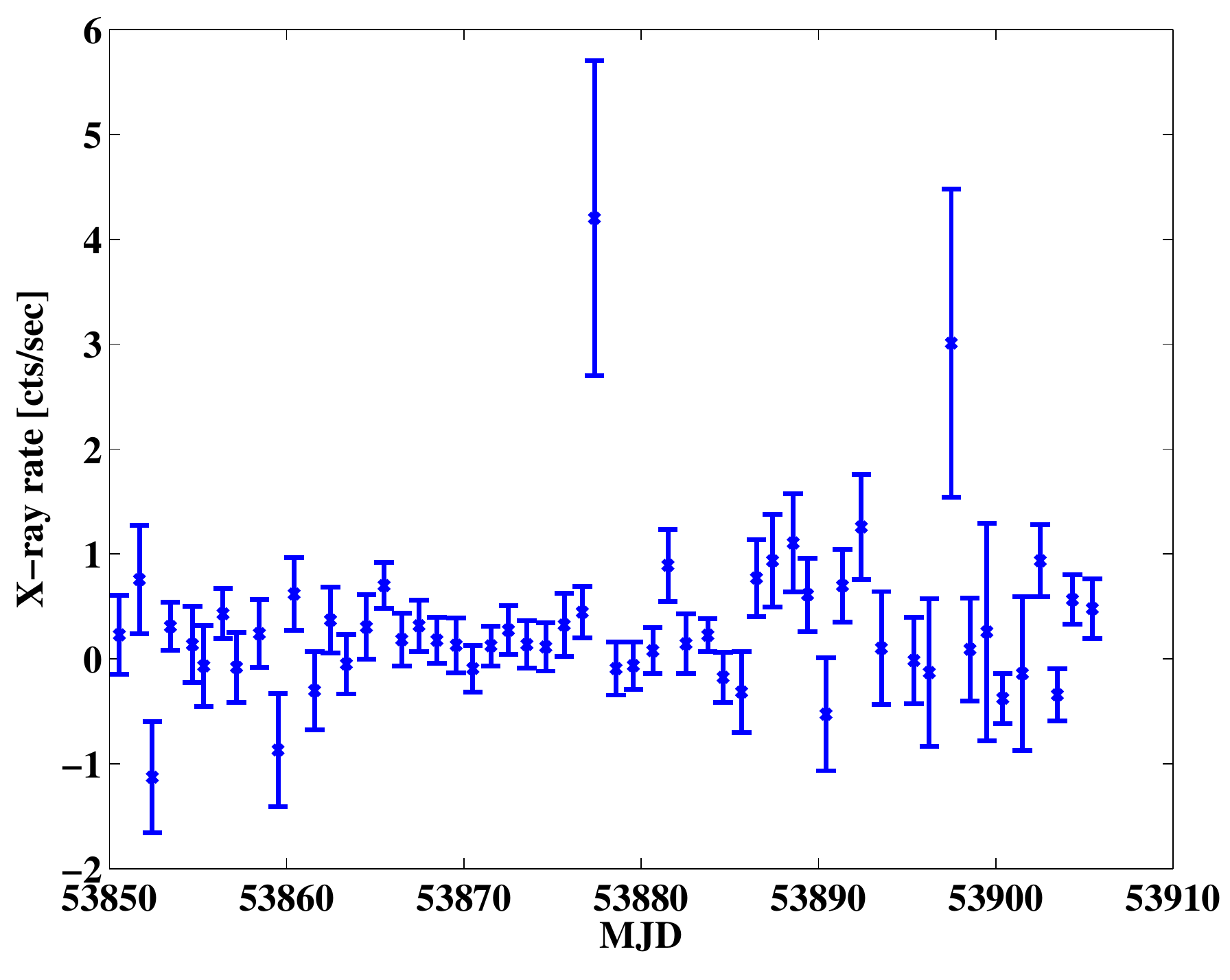}}
\end{center}
\caption{Left: rate of gamma rays detected from Mrk~501 with
VERITAS. Right: rate of X-rays detected by ASM on RXTEover the same
period}\label{FIG::501LIGHTCURVES}
\end{figure*}

During the period of the Mrk~501 observations it exhibited in a
considerably lower flux state than Mrk~421, but gave a consistently
positive signal for VERITAS. The Whipple 10m instrument observed the
source simultaneously but lacked the sensitivity to consistently
detect it. In addition the All Sky Monitor (ASM), a wide-field 2--10
keV X-ray instrument on the Rossi X-Ray Timing Explorer (RXTE) was
unable to detect emission from Mrk~501 during most of this period,
figure~\ref{FIG::501LIGHTCURVES}~(right).

\section{Conclusions}

We have presented the results of observations of Mrk~421 and Mrk~501
taken with the first two telescopes of VERITAS. During this period
Mrk~421 was in an active phase, and clearly detectable by
VERITAS. Mrk~501 was detectable at a lower flux level.  Although it is
premature to claim that the VERITAS observations during spring 2006
represent a detection of the baseline emission from Mrk~501, they
serve as an indication of the role that this sensitive gamma-ray
instrument will play in advancing our understanding of the emission
processes in blazars.

\section{Acknowledgments}

VERITAS is supported by grants from the U.S. Department of Energy, the
U.S. National Science Foundation and the Smithsonian Institution, by
NSERC in Canada, by PPARC in the U.K. and by Science Foundation
Ireland.

X-ray results provided by the ASM/RXTE teams at MIT and at the RXTE
SOF and GOF at NASA's GSFC.

\bibliography{icrc0234}
\bibliographystyle{plain}

\end{document}